\documentstyle[12pt,aaspp]{article}

\begin{document}

\doublespace

\title{THE EXTENDED NARROW LINE REGION OF 3C 299\altaffilmark{1}}

\author{Carlos Feinstein\altaffilmark{2}}
\affil{ Observatorio Astron\'{o}mico, Paseo del Bosque, 1900 La Plata, Argentina}
\author{F. Duccio Macchetto\altaffilmark{3}, Andr\'e R. Martel, William B. Sparks}
\affil{Space Telescope Institute, 3700 San Martin Drive, Baltimore, MD 21218}
\author{Patrick J. McCarthy}
\affil{The Observatories of the Carnegie Institution of Washington, 813 Santa Barbara Street, Pasadena, CA 91101}

\altaffiltext{1}{Based on observations made with the NASA/ESA {\it Hubble Space Telescope}, which is operated by the 
Association of Universities for Research in Astronomy, Inc, under NASA contract NAS 5-26555}
\altaffiltext{2}{Member of Carrera del Investigador Conicet, Argentina}
\altaffiltext{3}{On assignment from the Space Science Department of ESA}


\begin{abstract}

We present  results of  HST observations of the radio galaxy 3C 299.
The broad-band  F702W (R) and F555W (V) images (WFPC2/PC) show an 
elliptical galaxy, with a comet-like structure extending to the NE in the 
radio jet direction. 
The [OIII]$\lambda$5007 emission line map, shows a bi-conical structure 
centered on the nucleus, that overlaps  the structure found in the 
broad-band filters. 
The radio core coincides with the center of the bi-conical structure 
and the radio axes are aligned  with the direction of the cones. 
These data show  clear evidence of  a strong interaction 
between the radio jet and the NE morphology of the galaxy.

We show evidence that this NE region is an ENLR; the line-ratio diagnostics 
show that models involving  gas
shocked by the radio-jet plus ionization from a precursor HII region, 
produced itself by the ionizing photons of the postshocked gas on the 
preshocked gas provide a good 
match to the observations.  
                     
We investigate the spatial behavior of the 
ionizing parameter $U$, by determining the  [OIII]/[OII] line ratio 
which is sensitive to the change of the ionization parameter, and  trace 
its behavior over the ENLR along the radio jet direction. We find that 
 [OIII]/[OII] does not follow a simple  dilution model, but rather that 
it is approximately constant over a large range of distance from 
the nucleus thus requiring a local source of ionization which seems to
be  compatible with 
the shock models driven by the radio jet.

\end{abstract}


\doublespace

\section{Introduction}

Over the last few years  we have undertaken a systematic survey of extragalactic radio sources 
(de Koff et al. 1996; Martel et al. 1998a,b ; McCarthy et al. 1997)
 in the  3CR Catalogue (Bennett 1962a,b; Spinrad et al. 1985) using  WFPC2 on  HST. These 
data allow us to investigate the relationships between the radio and optical morphologies in a 
complete sample of powerful radio galaxies.
Imaging of nearby Seyfert galaxies with HST, such as Mrk 3, Mrk 6, Mrk 573, NGC 1068, NGC 2992 and NGC 4151, 
(Capetti et al., 1995a, 1995b, 1996, 1997; Winge et al. 1997;  Allen et al. 1998,1999; Axon et al, 1998) 
has shown an intimate connection between the radio structure and the extended NLR (ENLR). 
These studies show that the interaction of the radio jet with the ISM is the main  source of UV photons 
that ionize the ENLR in Seyfert galaxies.
We wish to investigate whether this scenario is also applicable to the high-power radio sources 
and in particular the radio galaxy 3C 299.

This galaxy is a FR II and one of the most asymmetric double-lobed radio AGNs (Liu et al. 1992)
not only because of the significantly different distance between the nucleus and each
of the  lobes but also in terms of lobe flux and lobe length. The  radio morphology consists of 
a compact radio-core and two lobes separated by 11$''$ at PA $\sim$ 63$^o$.
The NE lobe is larger ($\sim$ 1.5$''$) and brighter (441.6 mJy at 8.4 Ghz,
Akujor et al. 1995) than the SW lobe ($\sim$ 0.7$''$ and 20.1 mJy).
The core does not appear to be located at the geometrical center of the lobes; it is 3$''$ nearer 
the NE lobe (Liu et al 1991). For some time, due to this large asymmetry,
the NE lobe itself was believed  to
be the central component of a compact steep-spectrum source
(CSS). But when the SW component was found by Laing (1981), and the central
component by Liu \& Pooley (1991) and van Breugel et al (1992), the true
picture of 3C 299 as a double-lobed radio AGN fully emerged.

At optical wavelengths this galaxy appears to be a low-redshift analogue of 
the high redshift  (z$>$1) radio
galaxies studied by McCarthy et al. (1995) since its optical
continuum emission is elongated and aligned with the radio source axis. The emission
lines have a complex morphology and kinematics, with the [OIII]$\lambda$5007 line
offset from the continuum peak (McCarthy et al. 1996).

In this paper we discuss the complex morphology of 3C 299  shown by the 
high spatial HST resolution images  and we compare it  with the radio data. 
We investigate the physical conditions of the ENLR through ground-based 
long-slit spectroscopy data and show that the ionization parameter is 
roughly constant throughout the ENLR.

We conclude by showing that a simple mechanism to provide the required 
ionizing photons are shocks triggered by the interaction of the radio jet 
with the local ISM (Capetti et al. 1995a, 1995b), as in the case of the 
lower redshift Seyfert galaxies.

\section{Observations and data analysis}

The HST/WFPC2  observations of 3C 299 were taken as part of  the 
3CR Snapshot Survey (PI: W. B. Sparks) which was  conducted in cycles 4 to 7, 
in the F555W and F702W broad bands  as well as narrow emission line bands.
(de Koff et al. 1996; Martel et al. 1998a,b;  McCarthy et al. 1997).
The 3C 299 images were obtained with the WFPC2/PC and the F702W filter 
in May 1995 and the F555W filter in September 1996. Narrow-band images 
with the WFPC2/WF2 and the FR680N ramp filter were taken in August 1995. 
Table 1 shows the observation log.

At a redshift of z=0.367 , the galaxy
is at a distance of 1.3 Gpc with a projected linear scale of 4.7 kpcs/arcsec (assuming 
$H_{0}$=65 km sec${^{-1}}$ Mpc${^{-1}}$ and $q_{0}=0.5$).
The WFPC2/PC scale is 0$''$.0455/pixel and for the  WFPC2/WF2 the scale is 0$''$.0996/pixel,
and  therefore, the physical scales for the images are 213 parsecs/pixel for the PC mode and 
471 parsecs/pixel for the WF2 mode.

The reduction procedure for the F702W filter data, which is close to Cousins 
R filter, is fully discussed in Martel et al. (1998b). This reduction includes the 
standard WFPC2 pipeline processing and  cosmic-ray removal.
The data reduction was carried out using IRAF and the STSDAS package.
At the redshift of 3C 299, the F702W data includes the  
 H$\beta$ and [OIII] emission lines. The F555W filter data, 
which is  closer to Johnson V, was reduced in the same way 
and then registered  and added to produce one final image with higher 
S/N. Due to the redshift the F555W image includes the flux from the emission lines of
[OII], [NeIII], H$\delta$ and H$\gamma$. 
In what follows we will define F702W as R, and F555W as V.
The FR680N is a narrow-band ramp filter, which given the redshift of the galaxy and its
position on the CCD, produces an
[OIII] $\lambda$5007 image. These images were reduced as described above
and the final image was re-scaled to the resolution of the  WFPC2/PC.

The F702W and F555W filter data were flux-calibrated using values for the 
inverse sensitivities (PHOTFLAM) of 1.866 x 10$^{-18}$ ergs cm$^{-2}$ 
\AA$^{-1}$ DN$^{-1}$, 3.491410 x 10$^{-18}$ cm$^{-2}$ \AA$^{-1}$ DN$^{-1}$, 
and zero-points for the Vega system of  of M$_{Photzpt}$=-21.85, 
M$_{Photzpt}$=-21.07 respectively. Non-rotated images were used for these flux calibrations. 
The FR680N filter was calibrated using the ramp filter calculator 
(Biretta et al, 1996).
An approximate coordinate frame for the WFPC2 data is provided by the image
header information, based on the HST guide stars. 
The accuracy at the coordinates is approximately 1$''$ (Biretta et al. 1996).

Images obtained with different filters were registered to a common frame
using cross-correlation of the brighter elliptical structures. 
To derive the spatial shift between the broad-band filters images and the
FR680N image, we used the filamentary NE structure, since this structure
is well defined and bright in both  filters.
The sky background in each image was determined through the statistical
analysis of a 3-sigma clipping of the average background value 
in several regions in each of the images.

To compare with radio data, we used the observations from 
Leahy (1997), obtained  with MERLIN at 1534 MHz and posted on the DRAGN's 
Web site maintained by Leahy, Bridle, \& Strom R., at the Jodrell 
Bank Observatory.

\section{Results}

The optical broad-band images possess a complex morphology (Fig. 1). 
The V and R images show a central elliptical structure,
with two bright cusps separated by an absorption lane, $\sim$ 0$''$.3 x 1$''$.4 
(i.e. 1.2 x 5.7 kpcs).
To the NE, separated by 1.4 arcsec, there is an extended filamentary structure.
This region shows a shell-like structure with  bright, sharp 
but resolved filaments enclosing what appears to be a bubble of lower
intensity emission.
  
We found that the elliptical is well fitted by an $r^{1/4}$
surface brightness  profile. 
From the isophotal fitting, the position of the nucleus is found to be
hidden behind the central absorption lane.
From the V-R map, we derive values for the color index. The 
elliptical galaxy has a color index of  1.0 $<$ (V-R) $<$ 1.33,
which is compatible with the values from R\"onnbach et al. (1996) for the 
stellar population of a elliptical galaxy at the redshift of 3C 299. 
The faint region between the brighter cusps has 1.53 $<$ (V-R) $<$ 2.0 and
these values are also compatible with a stellar population reddened by a dust
lane.

The FR680N filter shows a rather different picture, since the data is 
dominated by emission in the [OIII]$\lambda$5007 line. The image (Fig. 2) 
shows a bi-conical structure of ionized oxygen, bright and large to the NE
and  faint and small to the SW. The NE cone has a full opening angle of 45
 degrees, from PA 30$^o$ to PA 75$^o$. 
The geometry and the brightness intensity (in comparison with the broad-band filters data)
 of this structure in the [OIII]$\lambda$5007 emission-line data indicates that 
this flux came from the  extended NLR (ENLR). This is consistent with the
previous studies of McCarthy et al. (1995, 1996).   
The shape of the NE bright lobe in this data is the
same as  that in the broad-band filters. Since there is almost no continuum
 emission (as the elliptical galaxy profile vanished) in the NE bright lobe, the 
broad-band filter morphology is dominated by the line-emission, therefore 
we used it to register the broad-band data to the narrow-band data.

Using a bi-dimensional cross-correlation technique (CROSSCOR task in IRAF/STSDAS) 
we find that the center of the bi-conical  structure coincides with the 
geometric center of the elliptical  galaxy. Fig 3 shows the superposition 
of the [OIII] data over the R data; note that the NE structure coincides 
with the  line emission. If we make the obvious assumption that 
the center (nucleus) of the elliptical galaxy is coincident with the radio core, 
it follows  that the radio axis (PA 63$^o$) is reasonably well aligned 
with the emission-line region (PA 52$^o$). Furthermore 
the NE radio lobe is clearly related to the ENLR in the [OIII] image and with the
large filamentary shell-like structure that is seen in the broad-band filters
(Fig. 4a,b).

We derived values for the physical conditions of the emitting gas by
analyzing and measuring line-emission ratios of the optical spectrum of
McCarthy et. al. (1996), taken with the Lick 3 m telescope.
The 2$''$ slit was positioned along the  radio source axis (P.A. $\sim$ 70$^o$),
thus crossing the  ENLR and covering a wavelength range from 4600  to 7400 \AA.
The McCarthy et al. data shows that 
the peak of the [OIII] and  H$\beta$ lines are offset from the continuum (see 
Fig 5 for a plot of this spectrum in the range from 6300 to 7400 \AA). 
Therefore the maximum of the line emission lies outside the nucleus 
and towards the ENLR. The high excitation line HeII $\lambda$ 4686 is present 
and shows a similar behavior with  extensions towards the ENLR.

One interesting point is to understand what 
dominates the flux in the broad-band images at the ENLR. For the F555W data the 
[OII]$\lambda$3727 and the [NeIII]$\lambda$3869 are present, and there is no observable continuum
in the ENLR spectrum. Since the [NeIII]$\lambda$3869 emission is 
low ($<$30\%) compared to the [OII]$\lambda$3727 and also
the [OII]$\lambda$3727/[NeIII]$\lambda$3869 ratio is quite constant over the entire ENLR,
it is possible to use the broad-band F555W image  as a good estimator of the [OII]$\lambda$3727 emission.  
We used  this image, to derive  surface brightness profiles and to determine  the
behavior of the [OIII]/[OII] ratio as an estimate of the spatial behavior of the  ionizing parameter ($U$). 
In Fig. 6 we show the [OIII] and [OII]  intensities as a function of distance from  the nucleus 
along the NE radio jet, as well as a plot of the [OIII]/[OII] ratio. The [OII] flux profile includes the
contribution from  the stellar population of the elliptical galaxy to a distance of
 about 1 arcsec from the nucleus, where the $r^{1/4}$ profile vanishes. Therefore, from 1$''$ to 3$''$.2, the 
measured flux arises  only from ENLR emission lines. 

The low amount of contamination of [NeIII] emission in the  [OII] data does not affect the validity of 
the [OIII]/[OII] ratio. Models such of  Dopita el al. 1995, show that the [NeIII] emission follows  the trend of the [OIII] 
emission as the effect of the precursor become more important. Sophisticated photoionization  models such as
those at Binnette et al. (1996)  show  a relation between the [NeIII] and [OIII] emission. Therefore, the  
properties of the
[OIII]/[OII] as the tracer of the ionization parameter $U$ are not affected by the low amount of [NeIII] emission.

The typical error for the [OIII] plot is $\sigma = \pm$ 0.7 for a reference flux 
of 4 $DN$ calculated with the Extend Exposure Time Calculator (Biretta et al. 1996)
and taking into account two pixel binning. For the [OII] plot, $\sigma = \pm$ 0.3 
for a 2 $DN$ flux and 8 pixel binning to match the spatial resolution of the upper plot 
(WF to PC resolution). The errors of the [OIII]/[OII] plot are a combination of the 
errors of the two upper plots and for example, a reference flux of 5 $DN$  has a 
maximum  error of $\sigma = \pm$ 0.9, calculated as the large departure
from the combination of fluxes and errors of the [OIII] and [OII] profiles.

\section{Discussion}

Our optical data shows that there is a close morphological relationship between the [OIII] ENLR and the radio emission of
the NE lobe. Is is interesting to investigate the possible physical scenario for this close relationship.
Taylor et al. (1992) proposed that fast bowshocks resulting from the interaction of the radio jet  and the ISM 
were the source of ionizing photons of the emission-line gas.
 Capetti et al. (1995a, 1995b, 1996, 1997) and  Winge et al. (1997) were the first to show that this mechanism 
best explains the optical  emission in the NLR 
 in nearby Seyfert galaxies (Mrk 3, Mrk 6, Mrk 573, NGC 1068, NGC 4151 and NGC 7319).
 Recently,  this work has been confirmed by Aoki et al. (1999) and Kukula et al. (1999).

Dopita \& Sutherland (1995a,b) have modeled in detail the ionization 
of the ENLR by  shocks. In one  scenario which has been shown to work
for  Seyfert galaxies, the radio jet interacts 
with the local interstellar medium and shocks the gas. In this scenario, the hot postshock plasma 
gas produces photons that can diffuse upstream and downstream of the jet. Photons 
diffusing upstream can encounter the preshocked gas and produce an extensive 
precursor HII region, while those traveling downstream will influence the ionization
 and temperature structure of the recombination of the shock. 

This scenario appears very attractive in explaining the morphology of the radio and optical data in 3C 299.
To check the validity of this interpretation, we derived values for the physical conditions of the
 emitting gas by measuring and analyzing the line-emission ratios from the optical spectrum of
 McCarthy et. al. (1996).
In Table 2 we list the spatial behavior of the He II $\lambda$ 4686/H$\beta$ and
 [OIII] $\lambda$5007/H$\beta$ ratios derived from that  spectrum.
Figure 7 is the plot of the measure line-ratio at 3.6$''$ (from the nucleus) over the 
models of Dopita et al. (1995a), it is clear that the model which includes 
a high-speed shock ($\sim$ 400 km/sec) plus a precursor HII region  
photoionization (produced by photons traveling upstream from the shock) 
fits the observations rather well over the entire ENLR. The values of 
Table 2 are also fairly constant (within the errors) showing that 
physical conditions are similar over the jet path.

Another way to check the shock  scenario is to explore the spatial behavior 
of the ionization  condition
of the gas by estimating  the ionization parameter $U$. This parameter
is defined as $U=Q/4\pi r^{2}cN_{e}$, where Q is the rate at which ionizing photons
are emitted from a point source towards a gas cloud with electron density $N_{e}$
and  distance $r$  from the ionizing source. 
If we consider the galactic nucleus to be the ionizing source, as the distance from the
nucleus increases for 
a constant $N_{e}$, the parameter U will decrease and change the
ionization condition of the gas. This change will be reflected in
the values of the  emission-line ratios (eg. [OIII]/[OII]). However, this
parameter will have a very different behavior  if the density changes 
or if there is another source of ionizing photons (such as shocks produced 
by the radio jet) or a combination of both (from the nucleus and
local).

The two brighter peaks in the  flux profiles  of both [OIII] and [OII],
 at 1.7$''$ and 2.4$''$ from the nucleus arise from the bright filaments 
of the shell-like structure. Note that both profiles have a similar shape. The [OIII]/[OII] 
line ratio is rather high  from 1.6$''$ to 2.6$''$ 
at the position of these filaments. As noted above this line ratio is 
strongly dependent on the ionization parameter $U$. If there is no local 
source of ionizing photons, due to
geometrical dilution of the ionizing radiation field we expect
$U \sim N^{-1}_{e}r^{-2}$. To produce a constant value for $U$ the
electron density should also follow an $r^{-2}$ law. 

Wellman et al. (1996) using radio data for 14 radio-galaxies and 8 radio-loud quasars 
obtained an estimate of the lobe magnetic field and the lobe speed propagation. From 
these parameters, they derived 
an estimate for the ambient gas density in the vicinity of the radio lobe, for the 
galaxies in their sample using the equation for ram pressure
confinement. They found that the ambient gas density can be fitted  with a modified King 
density profile.
In particular in our region of interest, within the inner 11 $h^{-1}$ kpc, the 
density still falls in the flat part of the density distribution, as the core radius of
this King profile is 50 (or 100 depending on the model) $h^{-1}$ kpc. 
From these, it seems extremely unlikely that the ambient density can vary as $r^{-2}$, 
as would be required to keep the ionization parameter constant 
in the case that the only source of ionizing photons is the nuclear source.
We therefore exclude that ionization by the nuclear source alone can be responsible 
for the observed ionization of the
ENLR lobe.

Similar physical properties of the ENLR, namely a rather constant value for $U$, has been
reported for some nearby Seyfert 2 galaxies. Capetti et al (1995a), Kukula et al. (1999) show a similar
result for Mrk 3, where the radio jets and the optical emissions are very 
 closely associated. 
 In Mrk 6, Capetti et al. (1995b)  show that there 
 is also evidence for a transverse ionization structure to the south where the
 radio jet and the emission lines are co-spatial. In the case of Mrk 573 
(Capetti et al. 1996), the
ratio [OIII]/[OII] indicates that the ionization parameter actually increases with 
radius. 
These examples provide  strong evidence for the scenario where  line-emitting gas in the ENLR is 
compressed by shocks and causes the line emission to be highly enhanced in the 
region in which this interaction occurs. 
We can also compare  3C 299 to NGC 1068, one of the
closest Seyfert 2 galaxies with a bright radio source and a prominent 
radio jet which terminates in a extended radio-lobe. For NGC 1068, Capetti et al. 
(1997) found
that the geometrical dilution model of the nuclear radiation field cannot
explain the high excitation core formed by the brightest knots of the NLR. 
 An increase (3-4 higher) of the ionization parameter is also observed at a 
distance of
$\sim$4$''$ from the nucleus, where the radio outflow changes its morphology
 from jet-like to lobe-like. This could be easily explained if there is a 
density increase where the jet enters the lobe, presumably due to the compression 
of the backflowing radio cocoon at the jet working surface.

In the case of 3C 299, morphology arguments i.e.  the extent and filamentary nature of the ENLR,
 the structure of the ENLR and its association to the radio-lobe, as well as physical arguments,
 i.e. the value of the emission-line ratios 
 HeII$\lambda$4686/H$\beta$ and [OIII]$\lambda$5007/H$\beta$
 compared to the models of Dopita et al. (1995a) and the constant value for the ionization parameter $U$, which
 does not follow the dilution of the ionizing radiation field for a central source 
of UV photons, strongly support the same scenario, namely that the ENLR is a product of the
 interaction of the radio-lobe with the ISM gas. 

In contrast to the nearby galaxies, 3C 299 has an intermediate redshift (z=0.367), 
and its ENLR is considerably larger ($\sim$ 12 Kpcs).
Best et al. (1996,1998,1999), have carried out deep spectroscopic observations of
a number of powerful radio galaxies with redshift $z \sim 1$  and 
show  that the passage of the radio jet induces shocks through the host galaxy
and plays a key role in producing the emission-line gas properties of these objects.
Thus the physical connection between the radio jet and the  ENLR
appears to be at work not only for nearby galaxies, but also for those  
at higher redshifts. This could provide an explanation for the ``alignment effect'' 
found by McCarthy et al. (1987), for  high-redshift radio galaxies.

\section{Summary}

We have shown  that the elliptical galaxy 3C 299, has a prominent dust lane
and bi-conical extended emission line morphology that is similar to those of
some lower redshift Seyfert galaxies.

We found that the NE radio radio-lobe  lies within  the 
shell-like structure of the ENLR, suggesting a physical connection between the jet and the ENLR.
We tested the scenario where the radio jet  compresses and shocks the interstellar gas 
thus producing the observed  morphology.  From the spectra of McCarthy et al. (1996) we compared the 
line ratio behavior of the ENLR with the models of Dopita et al. (1995a). 
From these models we found that the observed spectrum is compatible with gas
shocked by the radio jet plus ionization from a precursor HII region, produced by 
the ionizing photons on the preshocked gas. 
We also investigated the spatial behavior of the ionizing parameter $U$, by determining 
the  [OIII]/[OII] line-ratio which is
sensitive to the change of the ionization parameter, and we traced its behavior over
the ENLR along the radio jet direction. We found that the [OIII]/[OII] ratio does not 
follow a simple  dilution model, but rather that it is approximately constant 
over a large area (3 $''$ from the nucleus, $\sim$ 12 kpcs ) thus requiring a
local source of ionization, and we found that shock models driven by the radio jet, 
provide the necessary ionizing flux.

Therefore, because of morphology arguments (the NE lobe is locate at the ENLR), 
physical arguments (the emission-line ratios are consistent with a shock plus 
an HII precursor region) and the constant value of the ionization parameter (which
does not follow the dilution of the ionizing radiation field for a central 
source of photons), we conclude that the ENLR of 3C 299 is the result  of the 
interaction of the radio jet with the ISM gas.

\acknowledgments

C.F. acknowledges the support from the STScI visitor program. The Fundacion 
Antorchas (Argentina) provided partial financial support of this work. We are very 
grateful to Mark G. Allen for providing us 
the emission-line diagnostic models for a shock plus an HII region precursor.

\pagebreak

\paragraph{Figure Captions}

\paragraph{Fig 1a}  F702W (R) image of 3C 299. N is at the top, E is to the left.

\paragraph{Fig 1b}  F555W (V) image of 3C 299.

\paragraph{Fig 2 } [OIII]$\lambda$5007 image. The extended-narrow line
region shows the bi-conical structure. It is  centered on the
galaxy nucleus and is bright and extended to the NE.
 
\paragraph{Fig 3} Overlap between the [OIII] and the R filter images. Note that the 
NE structure  consists mostly of line emission.

\paragraph{Fig 4a} Overlap of the radio map at 1450 MHz with the
R image.

\paragraph{Fig 4b} Overlap of the radio map at 1450 MHz with the [OIII] image.

\paragraph{Fig 5} Long-slit spectrum taken by McCarthy et al. (1996). Note that 
the peaks of the emission lines are offset from the continuum. Wavelength is redshift
corrected.

\paragraph{Fig 6} Flux profiles ($DN$ units) of [OIII], [OII] and [OIII]/[OII], 
as a function of distance  from the nucleus along  the
 radio jet. See text for discussions of errors bars.

\paragraph{Fig 7} Plot of the emission-line ratios over the models of shock + precursor of Dopita et al. (1995).
Models are for shock velocities of 200,300 and 500 km/sec and  magnetic parameter $B/n^{1/2}$ of 0,1,2 and 4 
$\mu G \ cm^{3/2}$ respectively. 

\begin{table}
\caption[]{Log of observations}
\begin{flushleft}
\begin{tabular}{llll}
\hline
Filter Name & Em. Lines & Exp. time & Date   \\
 &   \ & sec &  \\
\hline

F702W &  & 300 &  May 1995 \\
FR680N & [OIII] & 300 & Aug. 1995  \\
FR680N & [OIII] & 300 &   Aug. 1995 \\
F555W &  & 300 & Sep. 1996 \\
F555W &  & 300 & Sep. 1996 \\
\hline
\hline
\end{tabular}
\end{flushleft}
\end{table}

\begin{table}
\caption[]{Emission-line ratio}
\begin{flushleft}
\begin{tabular}{lcc}
\hline
Distance from the nucleus & log(He II $\lambda$ 4686/H$\beta$) & log([OIII] $\lambda$5007/H$\beta$)   \\
arcsec &   &  \\
\hline
0& -0.57 $\pm$ 0.05 & 0.83 $\pm$ 0.02\\
0.7 & -0.60 $\pm$ 0.04 & 0.79 $\pm$ 0.02\\
1.5 & -0.62$\pm$ 0.04 & 0.90$\pm$ 0.02\\
2.2 & -0.66$\pm$ 0.05 & 0.87$\pm$ 0.02 \\
2.9 & -0.56 $\pm$ 0.06 & 0.79$\pm$ 0.03 \\
3.6  & -0.60 $\pm$ 0.08 & 0.83 $\pm$ 0.03 \\
\hline
\hline
\end{tabular}
\end{flushleft}
\end{table}

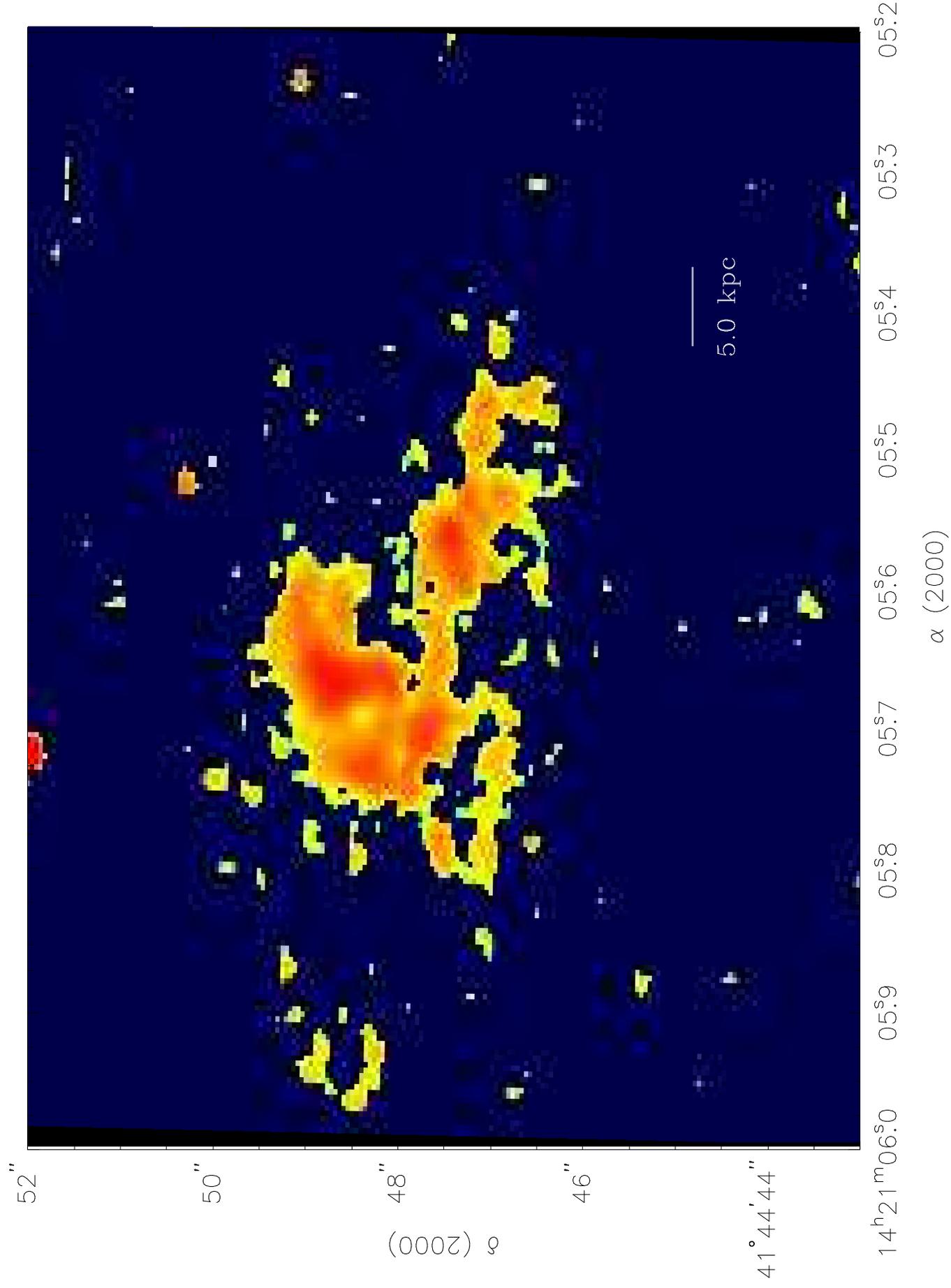
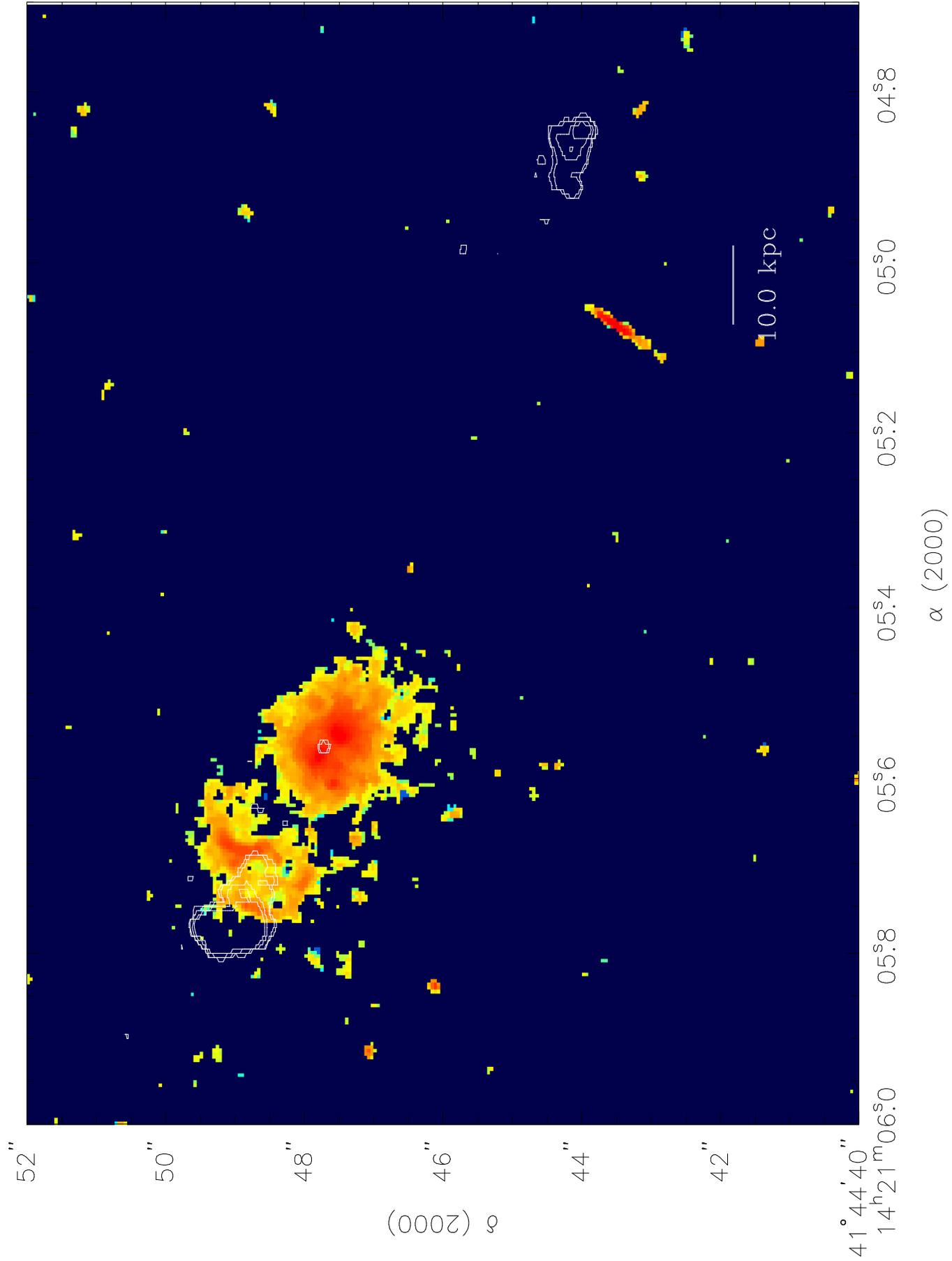
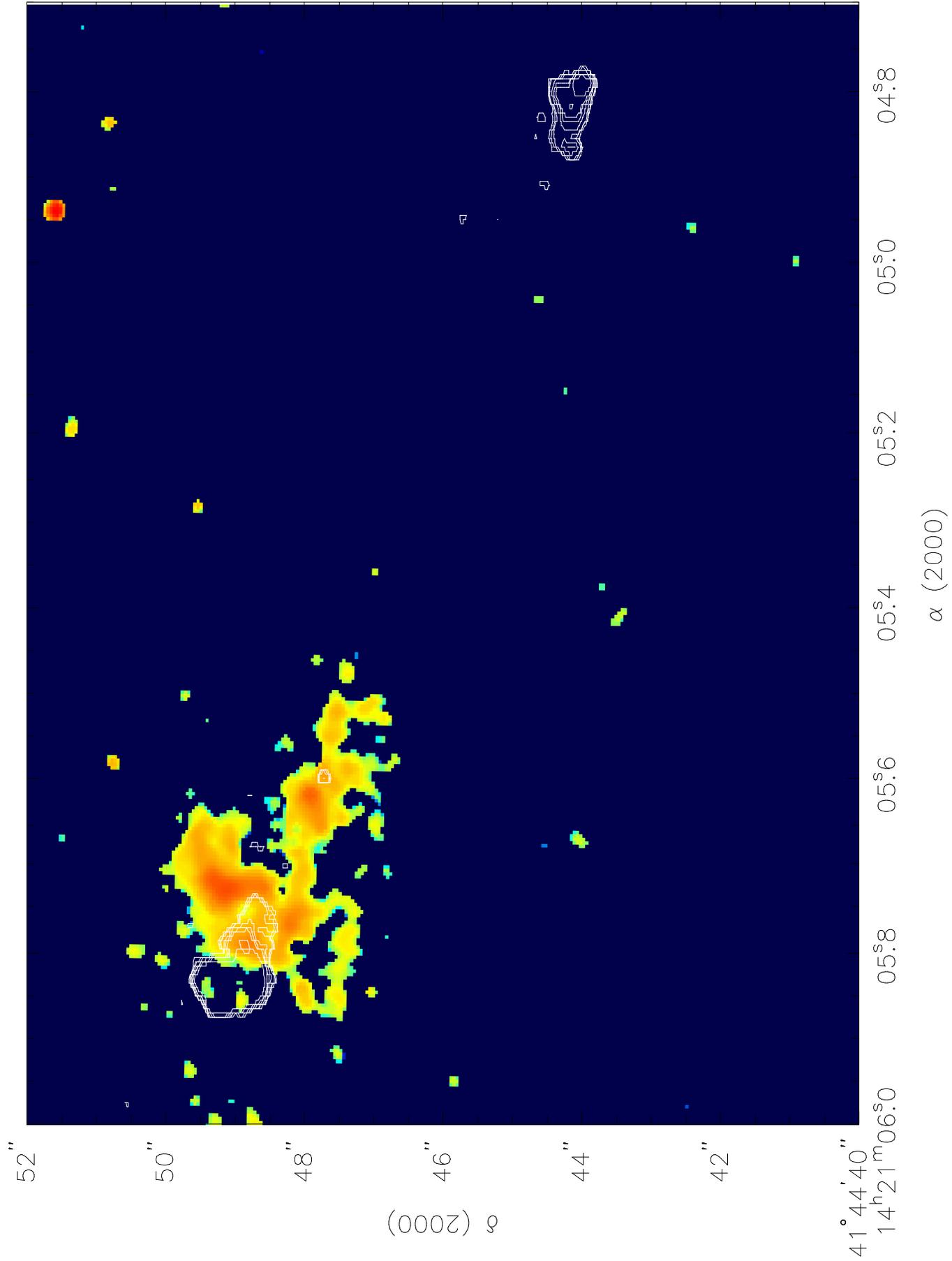
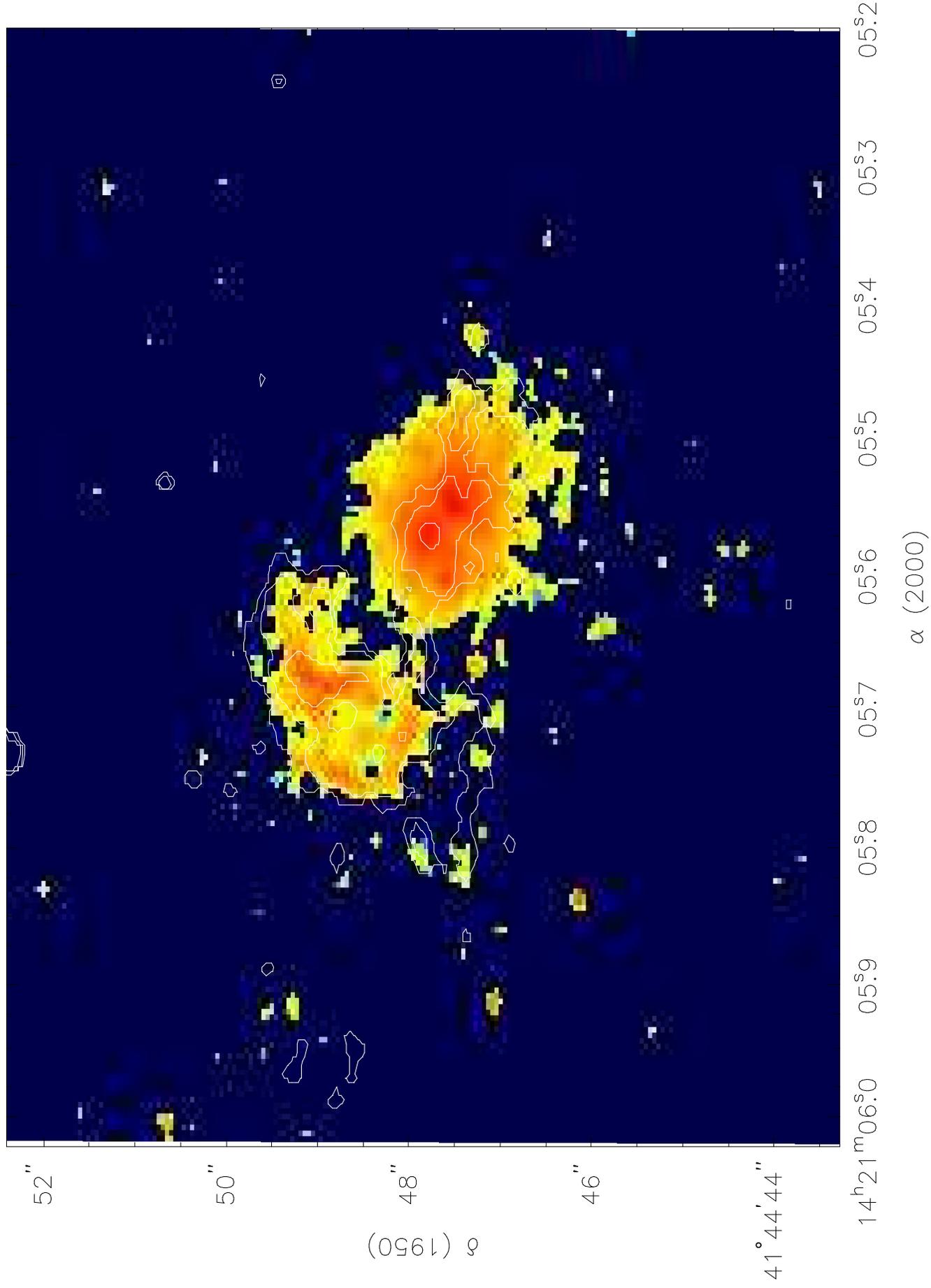
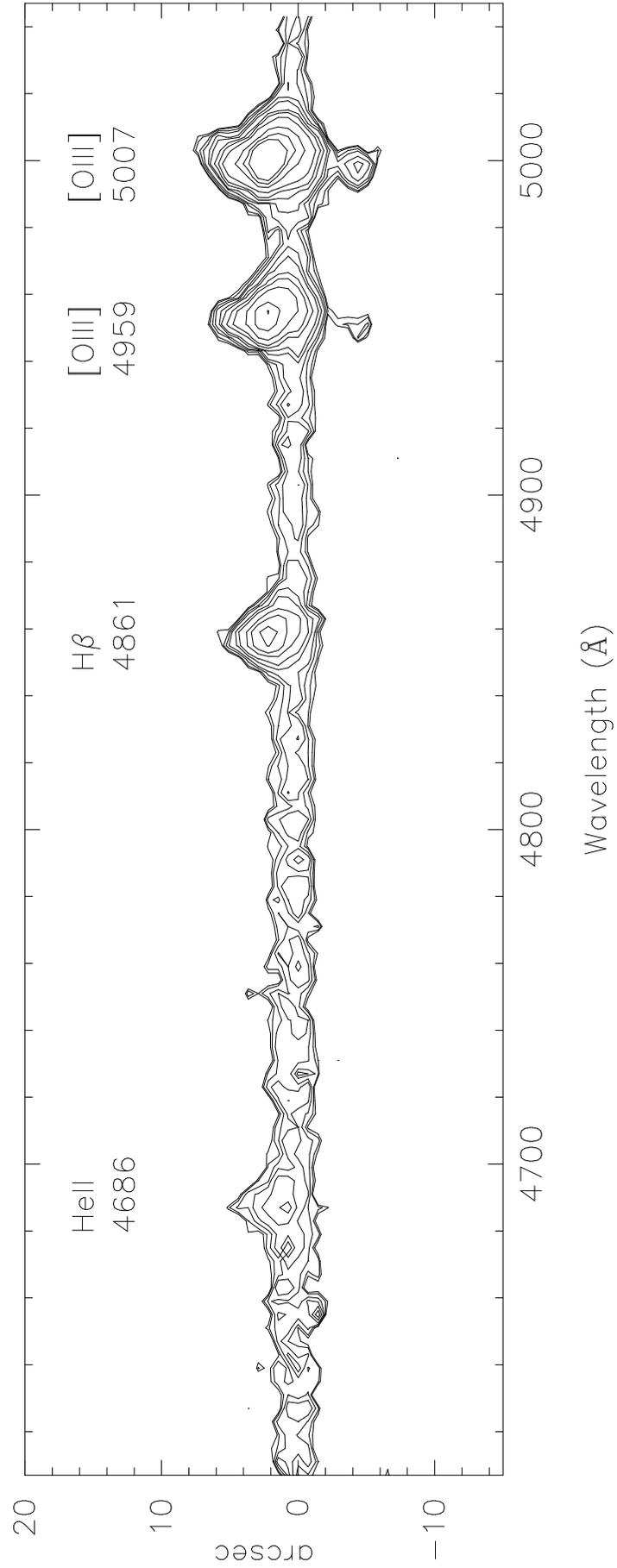
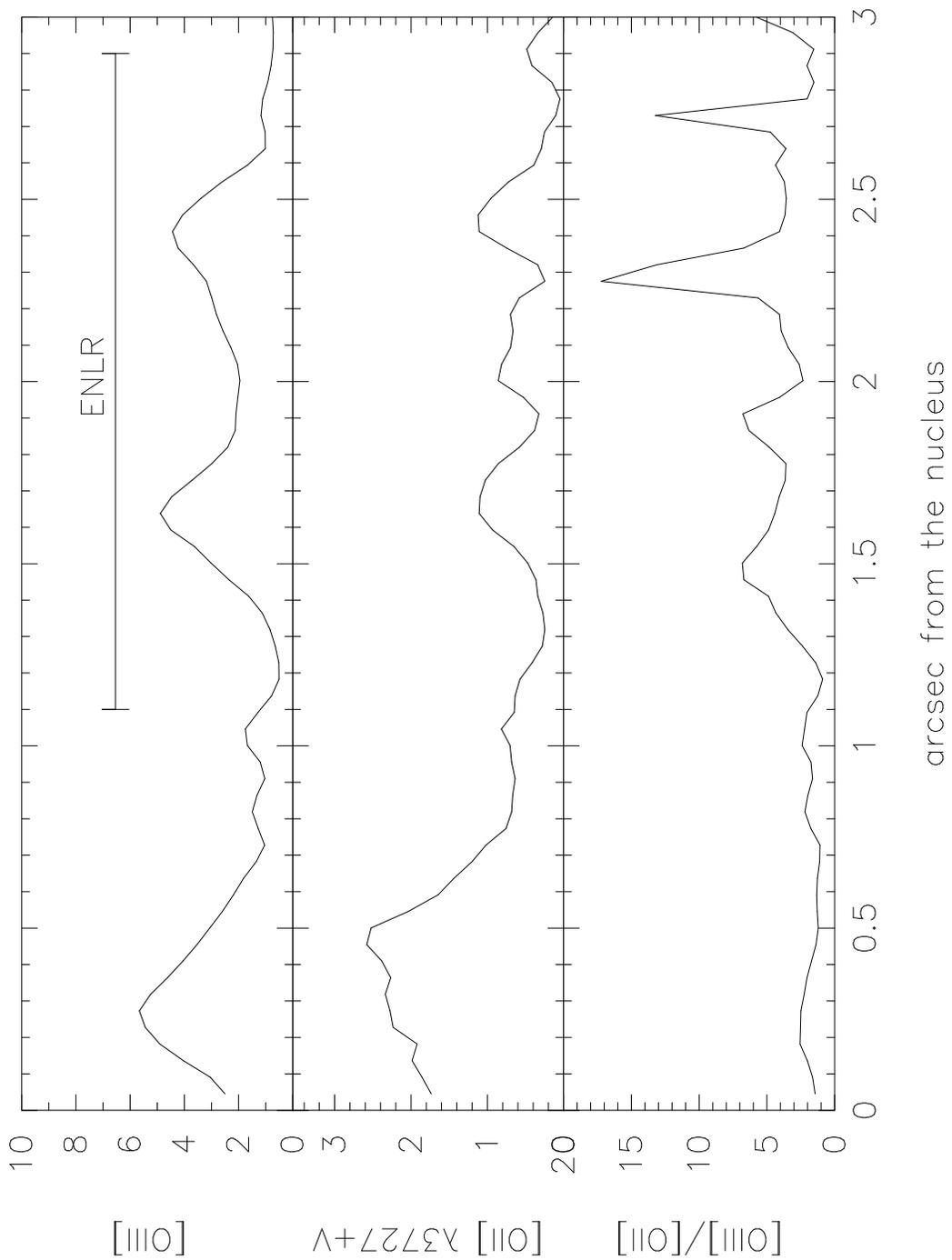
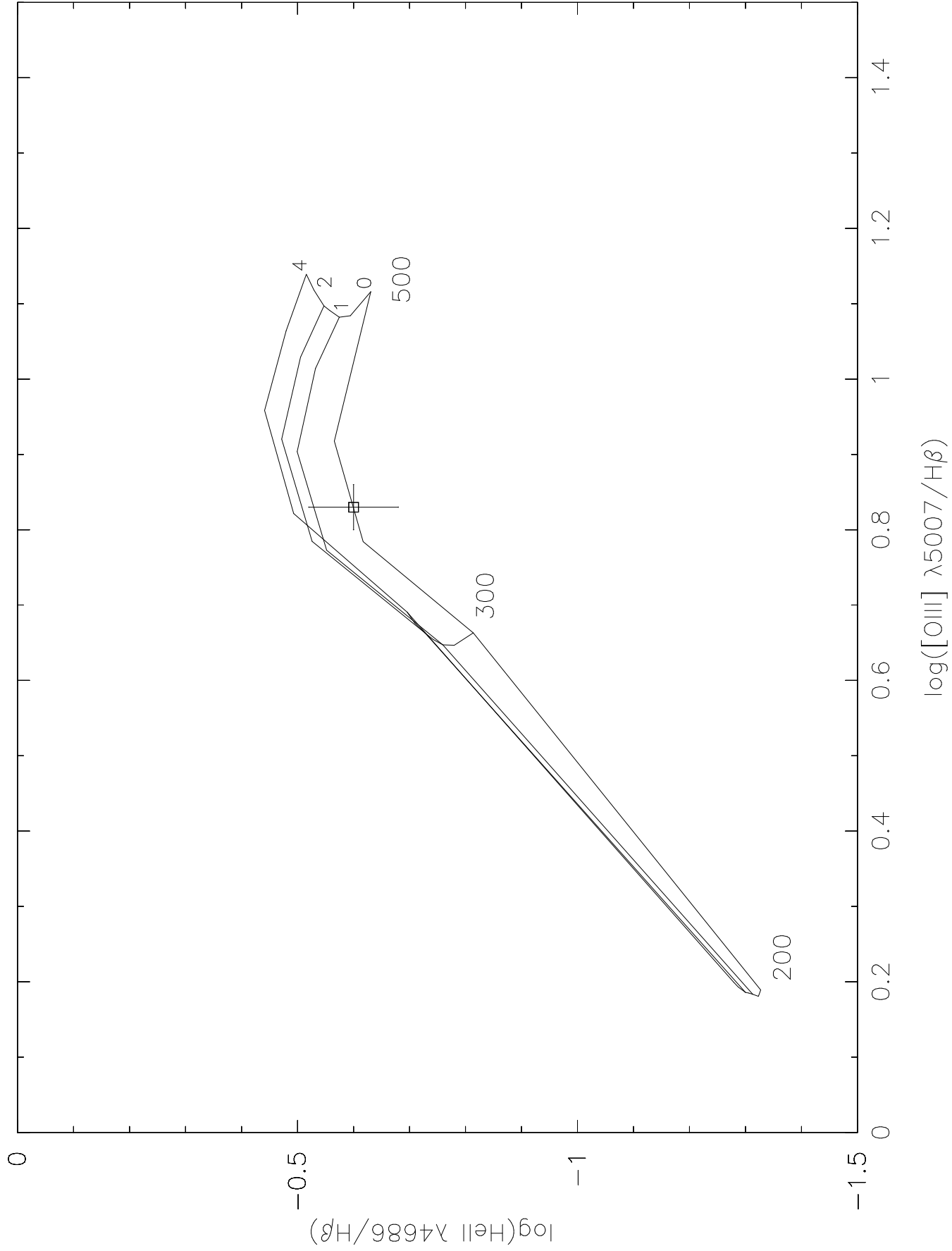
\end{document}